# A Novel Kalman Filter Based Shilling Attack Detection Algorithm

Xin Liu, Yingyuan Xiao*, Xu Jiao, Wenguang Zheng, Zihao Ling
*Tianjin Key Laboratory of Intelligence Computing and Novel Software Technology, Tianjin University of Technology, Tianjin 300384, China*
*Key Laboratory of Computer Vision and System, Ministry of Education, Tianjin University of Technology, Tianjin 300384, China*
liu_test@126.com, yyxiao@tjut.edu.cn, jiaoxu1999@sina.com, wenguangz@tjut.edu.cn, zihaoling@hotmail.com

***Abstract*—Collaborative filtering has been widely used in recommendation systems to recommend items that users might like. However, collaborative filtering based recommendation systems are vulnerable to shilling attacks. Malicious users tend to increase or decrease the recommended frequency of target items by injecting fake profiles. In this paper, we propose a Kalman filter-based attack detection model, which statistically analyzes the difference between the actual rating and the predicted rating calculated by this model to find the potential abnormal time period. The Kalman filter filters out suspicious ratings based on the abnormal time period and identifies suspicious users based on the source of these ratings. The experimental results show that our method performs much better detection performance for the shilling attack than the traditional methods.**

## I. INTRODUCTION

With the advent of the Big Data age, the recommendation system has become an important tool for users to choose potential projects of interest. The recommendation system can make recommendations for users based on the implicit connection between users and items. However, this system is relatively vulnerable to malicious users. The attacker deliberately inserts attack profiles into the recommender system to bias the predicted rating of a specific item. As a result, this item can be recommended to the influenced users more or less frequently. For example, if a user likes to watch comedy films and doesn't like to watch horror movies, there should be no horror movies in his recommendation list. After the attack, his recommendation list showed horror movies that he did not like. This user is called the influenced user, and this kind of attack is called “profile injection attack” or “shilling attack”.

In order to prevent shilling attacks, researchers tried to use some algorithms to detect attackers. The main challenge is how to detect attack profiles accurately and efficiently. The existing detection methods mainly use some features and models to separate suspicious users from the normal users. Specifically, some methods only consider the user’s rating of the item when finding attacker, and others consider more factors, such as time and user group. Although effective, the existing methods still have some limitations. Most of the detection algorithms are from the user's point of view, rarely from the perspective of the item, and the recall is poor.

According to the issues mentioned above, we explore a novel model for attack detection. We assume that an anomaly condition will occur when an attack occurs. In the case of sufficient data, the genuine user's rating behavior is basically stable. However, the attack profile injected by an attacker is usually done centrally in a certain period of time. Based on the hypothesis, we try to seek a time-related model to detect the unusual behavior and determine the attack profiles.

In this paper, we propose a novel Kalman filter model for detecting shilling attacks. In order to achieve the effect of detection, the main challenge is finding the attack profiles. According to the above analysis of the attack behavior, we use the Kalman filter's predicted value to compare with the actual value to calculate the deviation value. We define two kinds of deviation values and then determine the suspicious time period according to the abnormal condition of the deviation. Then the attack profiles are found from the suspicious time period. We compared the proposed Kalman filter with some existing methods. The experimental results show that our method performs much better detection performance for the shilling attack than the traditional methods.

The rest of paper is organized as follows. In Section II, we briefly introduce the related work. In Section III, we introduce some preliminaries, including attack profiles, models, etc. In Section IV, we describe our detection method in detail. In Section V, the results of the experiment are reported and analyzed. In the last section, we make a summary of the paper and prospect the direction of future work.

## II. RELATED WORK

Collaborative filtering is a popular recommendation algorithm which is very susceptible to shilling attacks. The attacker falsifies the user profile and make the fake users become close neighbors of genuine users as more as possible. Since collaborative filtering is recommended based on the interests of neighbors, the attacker can influence the system's recommendation results and increase or decrease the recommended frequency of the target object. There are already many researches on the detection methods of the shilling attacks in CFRS. Chirita et al. [1] proposed the RDMA attribute, which is the earliest defined attribute used to characterize the difference in user’s rating vector. Then develop an algorithm to calculate the probability of a user becoming a shilling attacker using RDMA and average similarity metrics. Burke et al. [2] present a number of detection features which are extracted from user profiles to classify users. They show that classifiers built using these features can detect attacks well and improve the stability of a recommender. Then, they utilize KNN model to complete the classification. Williams et al. [3] also extracted some features from user profiles. But they suffered misclassification and low precision. In addition, they tried the time interval detection scheme and present an empirical evaluation of the anomaly detection methods which can be quite successful in identifying items under attack and time periods in which attacks take place. But its robustness needs to be improved. Mehta et al. [4] provided an in-depth analysis of shilling profiles and put forward two unsupervised algorithms based on PLSA soft clustering and PSA variable selection. The algorithm has better detection performance, but it needs to know the size of the attack in advance, and the utility is not high. Peng et al. [5] proposed the "interest kurtosis coefficient" according to the concentration of interest of users, and used unsupervised detection methods to select effective detection indicators for different types of attack attacks, and then designed an unsupervised detection algorithm based on feature subsets. Zhang et al. [6] proposed a novel graph-based unsupervised learning detection algorithm, which transforms the problem of detecting attacks into the problem of searching for the largest subgraph. It is difficult for the algorithm to choose a suitable threshold and the performance is not good when the fill size is small. Wu et al. [7] proposed a semi-supervised detection algorithm, which combines the naive Bayesian classifiers and augmented expectation maximization base on several selected metrics. The algorithm is time consuming, low recall rate, and narrow applicable breadth. Lv et al. [8] proposed a detection method based on SVM and KNN algorithm, which can achieve good detection results when the mark data is small, but the accuracy and recall rate are relatively low when the fill size is small. Gao et al. [9] analyzed two common features and defined four types of items, then proposed a time intervals detection approach. They have achieved good results on four types of items, but the practicality is low. Yang et al. [10] find a mapping model between rating behavior and item distribution. They use 8 item attributes and 12 rating attributes to construct a corresponding relationship between ratings and items. Then use these presented features to construct a mapping model for detecting shilling attacks. The detection results demonstrate the outperformance of the proposed method, but cannot reach a full level when the attack size is small. Bhebe et al. [11] propose a combiner strategy that combines multiple classifiers in an effort to detect shilling attacks. The proposed Meta-Learning classifier have a nice detection performance, but also has a low recall rate at low attack sizes. Zhou et al. [12] propose a detection algorithm based on SVM and target item analysis method. The method has high precision but low recall. Conclusions that can be drawn from existing research, including the attacker usually rated the target project as the highest or lowest rating, and the injected attack files are usually injected in a short period of time. So, the average rating of the item will produce a significant deviation in a short time. If this abnormal deviation can be detected, then the time of the injection attack can be determined to determine the attack profile.

* Corresponding authors.
E-mail addresses: yyxiao@tjut.edu.cn

## III. PRELIMINARIES

### A. *Attack profile and attack model*

Attackers with different attack intentions will adopt different attack methods. Shilling attacks can be classified into two types: push attack and nuke attack [13]. In the push attack, the attacker will give target items the highest rating. Conversely, in the nuke attack, the attacker will give target items the lowest rating. The attack profile in a general attack model is shown in Table I.

TABLE I. THE GENERAL ATTACK PROFILE

| $I^S$ | | | $I^F$ | | | $I^\emptyset$ | | | |
|---|---|---|---|---|---|---|---|---|---|
| $i_1^S$ | … | $i_k^S$ | $i_1^F$ | … | $i_k^F$ | $i_1^\emptyset$ | … | $i_k^\emptyset$ | $i_t$ |
| $\delta(i_1^S)$ | | $\delta(i_k^S)$ | $\sigma(i_1^F)$ | | $\sigma(i_k^F)$ | *null* | | *null* | $\gamma(i_t)$ |

In an attack profile, $I^S$ represents the set of selected items. It is determined by function $\delta$. For some types of attacks, the selected item is not used. $I^F$ represents a collection of filler items which are randomly selected by the attacker is determined by function $\sigma$. $I^\emptyset$ represents the unrated items which are not rated by the attackers. In addition, $i_t$ is the target item that is promoted or demoted. Each attack profile has a target item. $i_t$ as determined by the function $\gamma$ is $r_{max}$ or $r_{min}$ depending on whether the attack is a push or a nuke attack.

In a single attack, the target item $i_t$ usually rated the highest rating (in the push attack) or the lowest rating (in the nuke attack). The choice of $I^S$ and $I^F$ is determined by the different types of attack models. There are three most commonly used models: random attack, average attack, and bandwagon attack [14].

- In random attack, the filler items $I^F$ are randomly chosen from the non-target items, and the rating value of each item is subject to a normal distribution with a mean rating value of the entire data set. The selected item is not used in this attack. The target item $i_t$ rating for $r_{max}$.
- Average attack is similar to random attack. The difference is that the distribution of the filler item ratings is determined by the normal distribution of the mean values of the ratings of each item $i$.
- In bandwagon attack, the popular items with the highest number of ratings are chosen as the selected items and rated at the highest rating value. Filler items are randomly selected in non-target items and are rated near the average of each item's rating like the average attack. The target items $i_t$ for all of these attacks are rated for the highest or lowest rating (depending on whether it is a push attack or a nuke attack).

### B. *Kalman Filter*

The Kalman filter algorithm estimates the value of the unknown variable for the next time period by analyzing the observed data over time. It can predict the next state based on the previous state (Assume that the input measured value and measurement noise are both normally distributed). Kalman filter can be represented by the following two equations.

$$x_{n_t} = F_t x_{n_{t-1}} + w_t \quad (1)$$

$$z_t = H_t x_{n_t} + v_t \quad (2)$$

Equation (1) is called state equation and indicates the internal state of this system. $x_{n_t}$ represents the state vector of time $t$ with a noise $w_t$, and $w_t$ is a normal distribution, $w_t \sim N(0, Q_t)$. $Q_t$ is a covariance matrix. $x_{n_{t-1}}$, the state vector of time $(t-1)$, is the vector of the previous state of $x_{n_t}$ and $F_t$ represents a matrix of time conversion of the system. Equation (2) is called observation equation. It can output the observation value $z_t$ according to the system state. $z_t$ is an observation vector with noise $v_t$, which is an observation noise vector that conforms to the normal distribution $v_t \sim N(0, R_t)$. $R_t$ is a covariance matrix. Then, $H_t$ is the mapping from the state vector to the observation vector.

Kalman filter has two steps: the prediction step and the update step. In the prediction step, it estimates the state at time $t$ according to the state at time $(t-1)$. Then in the update step, it estimates the state of a more accurate $t$-time state based on the observed value at time $t$. The two steps process are as follow.

Prediction step:

$$\hat{x}_{n_{t|t-1}} = F_t \hat{x}_{n_{t-1|t-1}} \quad (3)$$

$$P_{t|t-1} = F_t P_{t-1|t-1} F_t^T + Q_t \quad (4)$$

Update step:

$$Kg_t = \frac{P_{t|t-1} H_t^T}{H_t P_{t|t-1} H_t^T + R_t} \quad (5)$$

$$\hat{x}_{n_{t|t}} = \hat{x}_{n_{t|t-1}} + Kg_t(z_t - H_t \hat{x}_{n_{t|t-1}}) \quad (6)$$

$$P_{t|t} = (I - Kg_t H_t) P_{t|t-1} \quad (7)$$

$Kg_t$ represents Kalman gain. $\hat{x}_{n_{t-1|t-1}}$ and $P_{t|t-1}$ are the value of them at time $(t-1)$. $\hat{x}_{n_{t|t-1}}$ and $P_{t|t-1}$ are the predicted value they get in the prediction step. Then, after the correction of the update step, the correction values $\hat{x}_{n_{t|t}}$ and $P_{t|t}$ are obtained.

An application example of Kalman filtering in the recommendation system is proposed in [15]. They use the model to predict user preference vectors according to user features. Then, generate the recommendation list. They denote $F_t$ and $H_t$ as unit vectors $I$. Noise as a standard normal distribution. The two-step formula is update to:

$$\hat{x}_{n_{t|t-1}} = \hat{x}_{n_{t-1|t-1}} \quad (8)$$

$$P_{t|t-1} = P_{t-1|t-1} + I \quad (9)$$

$$Kg_t = \frac{P_{t|t-1}}{P_{t|t-1} + I} \quad (10)$$

$$\hat{x}_{n_{t|t}} = \hat{x}_{n_{t|t-1}} + Kg_t(z_t - \hat{x}_{n_{t|t-1}}) \quad (11)$$

$$P_{t|t} = (I - Kg_t) P_{t|t-1} \quad (12)$$

We improved on this model and further transformed this model with the characteristics of the shilling attack.

### C. *Confidence interval estimate*

Confidence interval estimation is based on the central limit theorem which shows that the distribution of sample mean becomes more standardized as the sample size increases. So, if we have a large enough sample, we can use a normal distribution to describe the sample mean from any group, even a non-normal population [16]. The confidence interval or interval estimation is based on the point estimate and the sampling standard error, and the interval containing the parameter to be estimated is established according to the given probability value. The given probability value is called the confidence or confidence level. For example, a 95% confidence interval indicates that 95% of the data in a given data will fall within this interval. The two values delineating the confidence interval are called the lower confidence limit and the upper confidence limit.

Suppose we have some ratings $x_1, x_2 \ldots x_k$. The purpose of interval estimation is to find two statistics $y_1, y_2$, so that $(y_1, y_2)$ can cover these samples as much as possible. $y_1$ and $y_2$ can be calculated by the following formula.

$$y_1 = \bar{x} - \sigma Z_\alpha \quad (13)$$

$$y_2 = \bar{x} + \sigma Z_\alpha \quad (14)$$

Where $\bar{x}$ and $\sigma$ is the mean and standard deviation of these ratings. $Z_\alpha$ is the z-value at the confidence level $(1-\alpha)$. For example, when the confidence level is 95%, the value of $Z$ is 1.96.

### D. *Item classification*

In order to target the characteristics of different items, in [9], they are classified into four categories according to the z-score of the items and the average number of ratings, fad item, fashion item, style item and scallop item. Z-score is calculated from the life cycle of each item. The life cycle represents the difference between the start time $t_S$ and the end time $t_E$ of a time span. Here is the formula for z-score:

$$\mathrm{z-score}(x) = \frac{x - \bar{A}}{\sigma_A} \quad (15)$$

Where $x$ is a life cycle of an item; $\bar{A}$ and $\sigma_A$ are the mean and standard deviation of the life cycle of all these items, respectively. The specific division rules of the items are shown in Table II.

TABLE II. ITEM DIVISION RULE

| **Category** | **z-score (*z*)** | **Number of ratings (*n*)** |
|---|---|---|
| Fad item | $z < 0$ | $n \le \bar{n}$ |
| Fashion item | $z < 0$ | $n > \bar{n}$ |
| Style item | $z \ge 0$ | $n \le \bar{n}$ |
| Scallop item | $z \ge 0$ | $n > \bar{n}$ |

## IV. OUR APPROACH

### A. *Problem Analysis*

We see the shilling attack as an anomaly detection problem, then we need to find the exception. It is mentioned in [16] that the normal behavior of the recommendation system can be characterized by a series of observations over time. In other words, the rating in the normal recommendation system is regular. When an abnormality occurs, the regular rating will be disturbed. The biggest challenge we face is to find the potential unusual time period when the normal ratings became abnormal. Then we propose a detection algorithm based on Kalman filter model for the characteristics that the attack profiles are injected in a short time.

Prediction step:

$$\hat{x} = \frac{x(n_A+n_P)}{n_A} \quad (16)$$

$$\hat{P} = P + q \quad (17)$$

In the prediction step, first, we get the preliminary prediction value of rating sum *x* (the sum of the values of all ratings up to this state) and the model's standard deviation *P*. Then, $n_A$ represents the total number of ratings that have been calculated before the start of this condition. $n_P$ represents the total number of ratings in the current time block. $q$ represents the standard deviation of model error. Here, $[(n_A + n_P)/n_P]$ is equivalent to $F_t$ in (1).

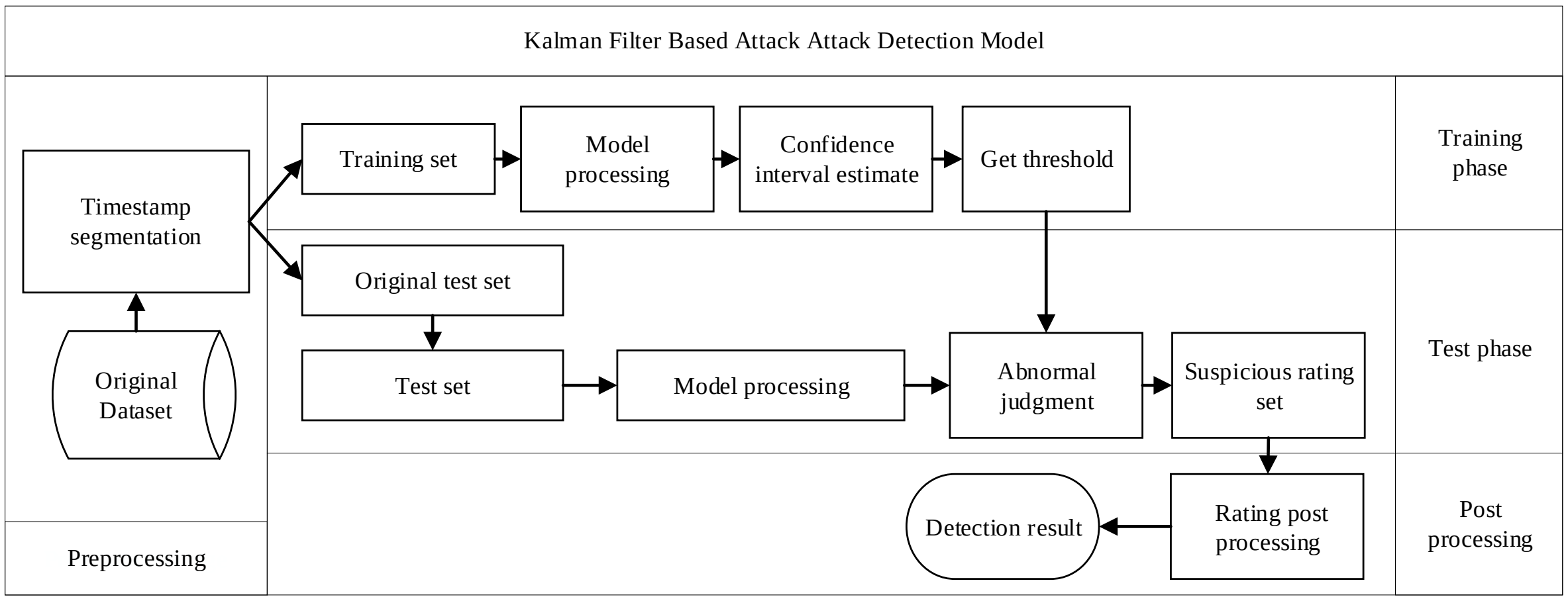

Fig. 1. The process of detection model

Update step:

$$Kg = \frac{\hat{P}}{\hat{P}+r} \quad (18)$$

$$x^+ = \hat{x} + Kg(z - \hat{x}) \quad (19)$$

$$P^+ = (1 - Kg)\hat{P} \quad (20)$$

$$n_A = n_A + n_P \quad (21)$$

In the update step, *x* and *P* and preliminary prediction values $\hat{x}$ and $\hat{P}$ are further processed to obtain more accurate prediction values $x^+$ and $P^+$ as output prediction results. In (18), $Kg$ represents Kalman gain and r represents the standard deviation of measurement error. In (19), *z* is the sum of the values of all ratings within the current time block (observation value). Equation (21) is to update the value of $n_A$.

In [15], they assume that user preferences do not change due to purchase actions, treating $F_t$ and $H_t$ in (1) and (2) as *I* (a unit vector), and taking into account the shift in user preferences, $w_t$ and $v_t$ in (1) and (2) are regarded as the standard normal distribution $N{\sim}(0, I)$. So here, we assume that *q* and *r* in (17) and (18) are 1.

### B. *Detection process*

The process of detection is as shown in Figure 1.

The detection process is divided into four phases, pre-processing phase, training phase, test phase and post-processing phase.

In the training phase, our goal is to get the threshold for detection. First, the training set extracted from the original data set is processed using our proposed model, which is described in the *C* subsection of our approach. Predicted values are generated during each step of the model. Then, the deviation is calculated from the comparison of the predicted value with the actual value. The section on deviations is described in the *D* subsection of our approach. The deviations produced at each step of the model are then counted together as a training set for the deviation. Interval estimation of these deviations produces a threshold. The choice of threshold is described in the *E* subsection of our approach.

In the test phase, our goal is to get the time period in which the anomaly rating occurs. We add part of the data from the original set to the attack profile as a test set, and generate the deviation test set after the same processing as the training set. Then, the deviation test set is screened according to the threshold obtained in the training step to obtain a suspicious rating set. The screening method is described in the *D* subsection of our approach. Finally, the suspicious rating set is post-processed to obtain the detection result. The post-processing process is described in the *F* subsection of our approach.

### C. *Detection model*

In this paper, we propose a Rating Detection Adapted Kalman Filter model (RDAKF). We have made some improvements to the traditional Kalman filter model.

### D. *Deviation scheme*

A recommendation system is basically stable when the amount of data is large enough. The deviation of the data is kept within a certain range. First, we declare the meaning of the deviation mentioned in this article here. The deviation mentioned in this paper refers to the difference between the value predicted by the model and the actual value in the data set.

The key point in judging anomalies is to determine whether the deviation is abnormal. So, we propose two deviations.

The total deviation:

$$v = y - \hat{x} \text{ where } y = x + z \tag{22}$$

The average deviation:

$$v_A = v / n_P \tag{23}$$

$y$ in the (22) represents the sum of the results of the previous step and the observed data of this step. In (23), $v_A$ represents how much each rating in this step deviates from this step.

With these two parameters, we can determine which deviations are abnormal. There are four possible situations.

- Both $v$ and $v_A$ are normal, indicating that the system has not been attacked at this time.
- $v$ is normal but $v_A$ is abnormal, that is, the total deviation is normal and the average deviation is abnormal. In this case, it is possible that the rating data is small (the total deviation is normal) during this time period, and both are extreme rating values (average deviation abnormality). This situation is normal and cannot be judged as abnormal.
- $v$ is abnormal and $v_A$ is normal. The average deviation at this moment is normal and the total deviation is abnormal. This phenomenon occurs at the time of the rating set, the number of ratings is large (the total deviation is abnormal), and both are normal rating (average deviation is normal).
- Both $v$ and $v_A$ are abnormal. There are many ratings and extreme ratings, and we believe that this is the time of attack.

In summary, when both $v$ and $v_A$ are abnormal, it is determined that the system is under attack.

### E. Threshold selection

In order to judge whether the deviation exceeds the standard, it is necessary to set a threshold. To this end, we combine the interval estimation knowledge in the *C* subsection of Section III to select a reasonable threshold.

We treat each deviation as a sample of data. Confidence interval estimates for individual samples were performed on these samples. Choose a suitable confidence level to get a confidence upper bound. And use this upper bound as the critical value, which is the threshold, as the basis for judging the abnormality.

Two thresholds are needed to determine the anomaly of two deviations, and these two thresholds need to be trained from the training set. Two upper bounds of confidence intervals $\eta$ and $\eta_A$ are obtained as thresholds. If $v > \eta$ and $v_A > \eta_A$, it is regarded as an abnormal point.

### F. Post-processing

The detection effect of the model can only be locked to the time period during which the shilling attack occurs, and the ratings within the time period need to be further filtered. When we get a series of ratings for a suspicious time period, we need to remove the ratings that are not extreme ratings (if it is a push attack, the rating that is not the highest rating is removed, and if it is a nuke attack, the non-minimum rating is removed). For a shilling attack, if attacker do not use extreme rating to attack the system, the effect of the attack will not easy to achieve the ideal situation of the attacker. Then find users who rated these extreme ratings and convert the suspicious rating set into a suspicious user set as the final detection result.

## V. Experiments

### A. Experiment setup

In the experiment, we use the 100K Movielens dataset, which contains 100,000 ratings from 943 users and 1682 items, records a total of 215 days of data from September 20, 1997 to April 23, 1998. Each user in this dataset rates at least 20 movies. The rating ranges from 1 to 5. We treat all the data in the original data set as the data evaluated by genuine users and divide this data into parts every four days, for a total of 54 parts. We used 100 items including 5328 deviations as training set for training. According to the proportion of the item types, 12 of them are fad items, 6 of them are fashion items, 32 of them are style items, and 50 of them are scallop items. Total deviation selects 99% confidence and average deviation selects 90% confidence.

This experiment is all about push attack. We select 5 projects as target profiles. Insert different scales of attack size with filler size 5%, and make them the same size as the genuine users. For random attacks and average attacks, we generate ratings from $N \sim (\bar{r}, \bar{\sigma}^2)$ and $N \sim (\bar{r_i}, \bar{\sigma_i}^2)$. In the bandwagon attack, we select movie 50 as the selected item and rate it $r_{max} = 5$.

We tested our method on three attacks and then compared it with KNN, Bayes[11] and SVM[12] algorithms. KNN is a traditional classification algorithm. It is the grouping of the k most similar users. Bayesian detection algorithm mentioned in [11] use the combiner strategy that combines multiple classifiers in an effort to detect shilling attacks. The SVM algorithm in [12] combines target item analysis method on the basis of the SVM model to improve the detection performance.

### B. Evaluation Metrics

To measure the performance of the detection algorithm, we use two metrics for precision and recall.

$$Precision = \frac{TP}{TP+FP} \qquad Recall = \frac{TP}{TP+FN} \tag{24}$$

Precision reflects the proportion of a category of targets detected by the detector that really belong to that category. Where *TP* is the number of attack profiles detected correctly, *FP* is the number of misclassified genuine profiles into attack profiles.

Recall reflects the true number of categories in a category of targets detected by the detector. *FN* is the number of attack profiles misclassified into genuine profiles.

## C. Experimental Results and Discussion

In this section, we test the performance of our detection algorithm. First, we test the performance of our algorithm under different attack models. Then, we test the impact of different confidence levels on detection performance. Finally, we perform performance comparisons with the other three algorithms.

### 1) Performance of Different Attack Models

We use our algorithm to test its precision and recall in dealing with random attack, average attack and bandwagon attack. The filler size is set to 5%.

Figure 2 shows the precision of our approach under different attack sizes. It can be seen that as the size of the attack increases, the precision of the detection algorithm is increased and the precision remains above 0.8. However, compared to the other two attacks, the algorithm has a poor effect on the low attack size of bandwagon attack. It can be seen that our algorithm has more misclassification in bandwagon attack. Bandwagon attack contains selected item, which is not included in the other two types of attacks. From the experiment result, we can conclude that the selected item has a certain impact on the detection performance of our algorithm.

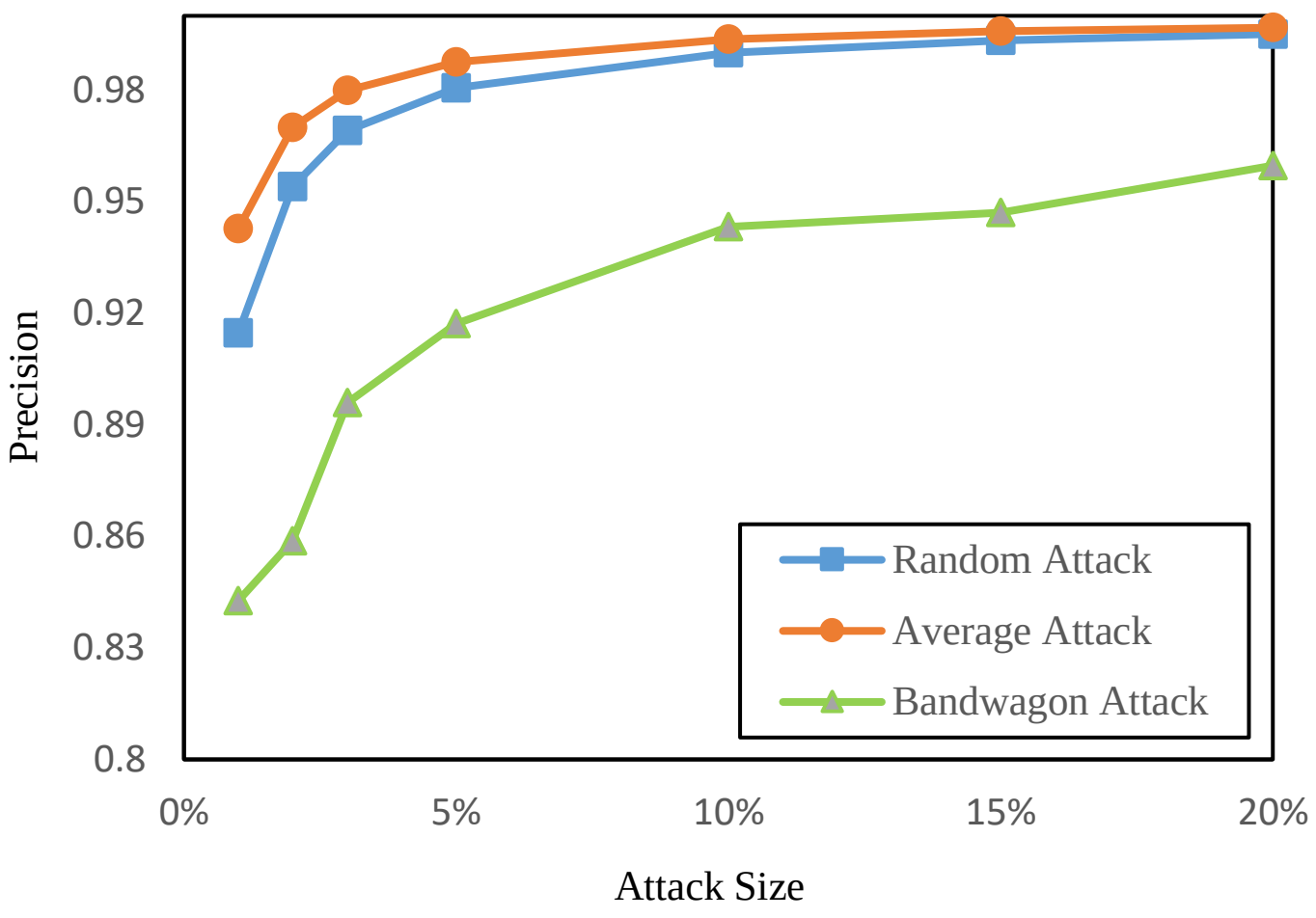

Fig. 2. Precision of different attack models when the attack size varies and filler size is 5%

Then, Figure 3 is the recall of our method under different attack sizes. Our algorithm has the best detection rate for average attack, and it is acceptable for random attack, and the same problem with precision is not effective for bandwagon attack. It can be seen that the selected item also has an impact on the recall.

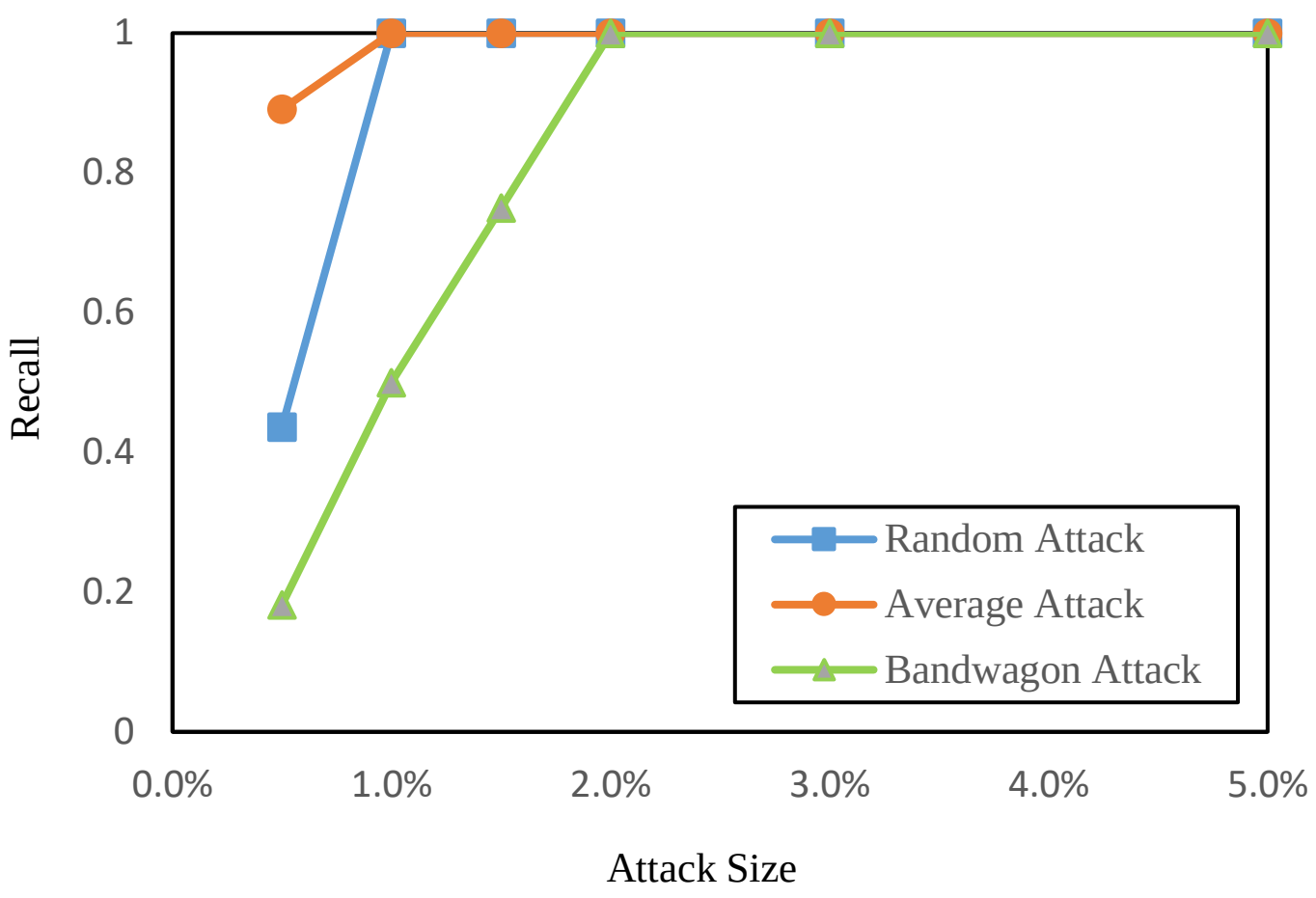

Fig. 3. Recall of different attack models when the attack size varies and filler size is 5%

### 2) Impact of Confidence Levels

The choice of confidence level will affect the precision and recall. If the confidence level is high, then the range it accepts will expand. It means that a larger deviation can be accepted. The deviation caused by some attacks may not be detected, that is, the recall is reduced. However, due to the increase in acceptability, the probability of a normal deviation being misjudged is reduced, that is, the precision is improved. Conversely, if the confidence level is low, the recall will increase and the precision will decrease. So we compare the impact of different confidence level selections on performance to choose an appropriate one.

Figure 4 shows the detection effect of the confidence levels for the different total deviations when the confidence level of the average deviation is 90%. It can be seen that the higher the confidence level of the total deviation, the higher the accuracy when the confidence level of the average deviation is fixed.

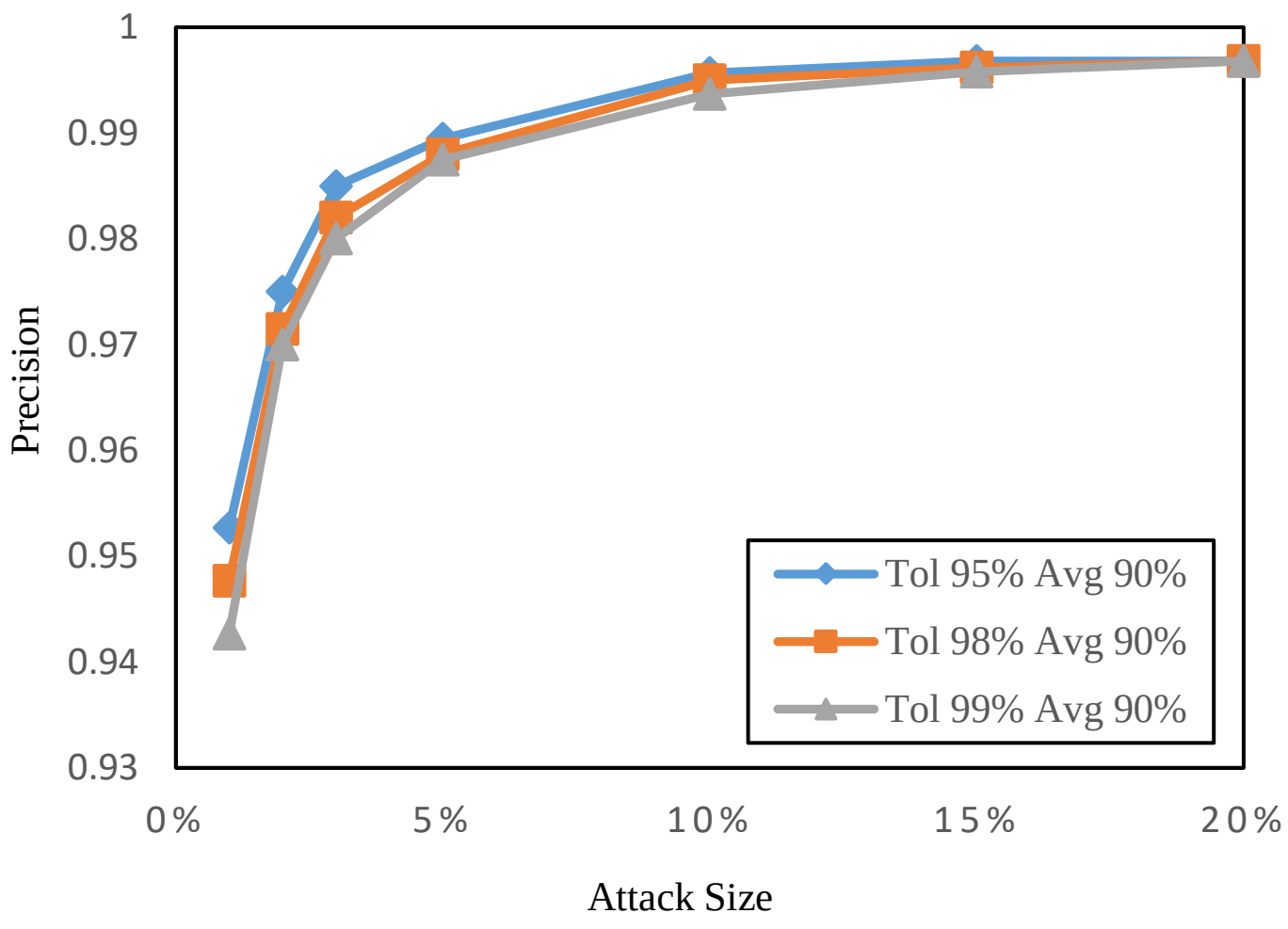

Fig. 4. The effect of different confidence levels of total deviation on detection

Figure 5 and 6 show the detection effect of the confidence level of the total deviation and the confidence level of the different average deviations. Figure 5 shows the average attack and Figure 6 shows the bandwagon attack. In average attack, 90% confidence level and 85% confidence level are comparable, while 95% confidence level is less effective. In bandwagon attack, 95% confidence level highest, 90% second, 85% worst.

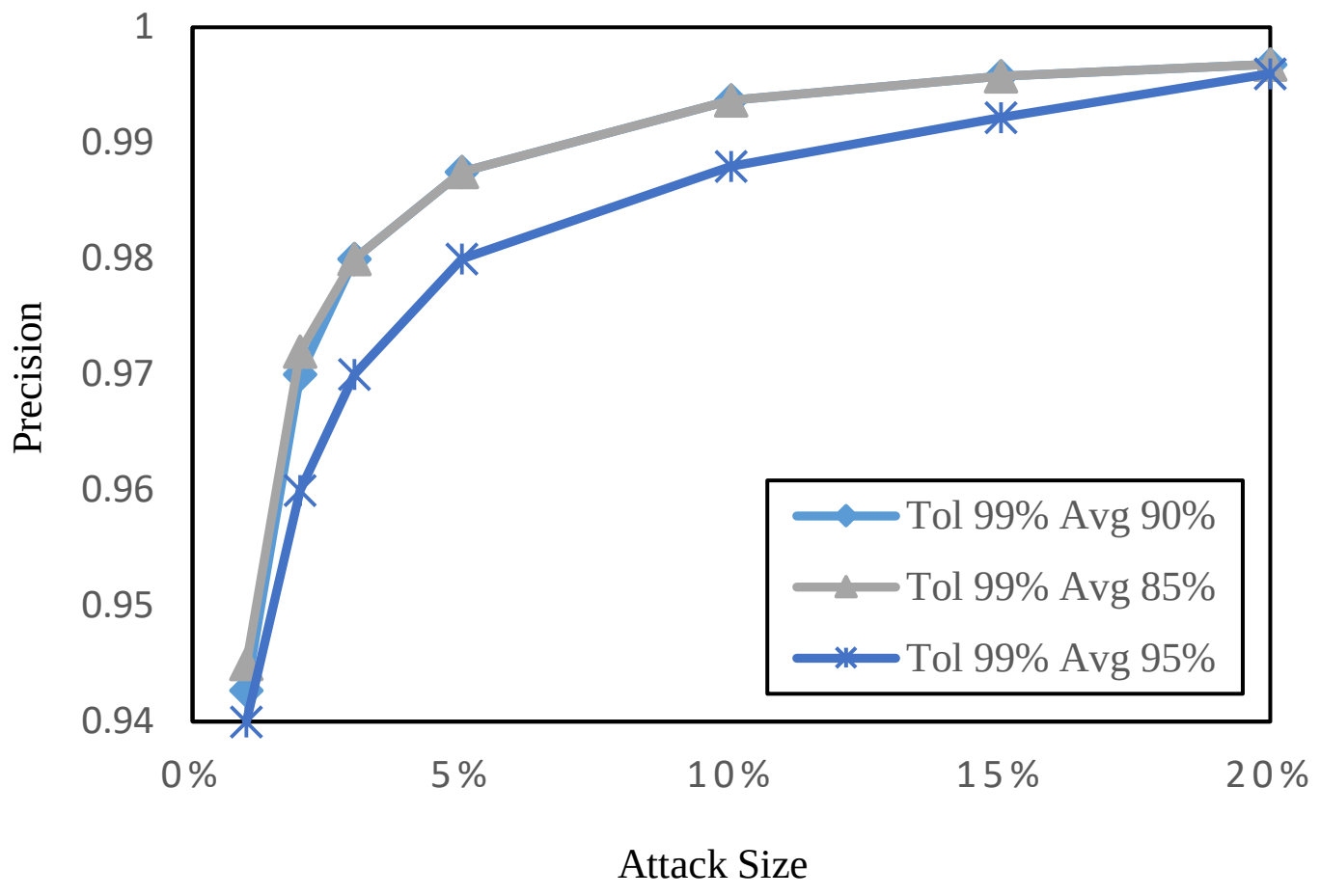

Fig. 5. The effect of different confidence levels of average deviation on detection (average attack)

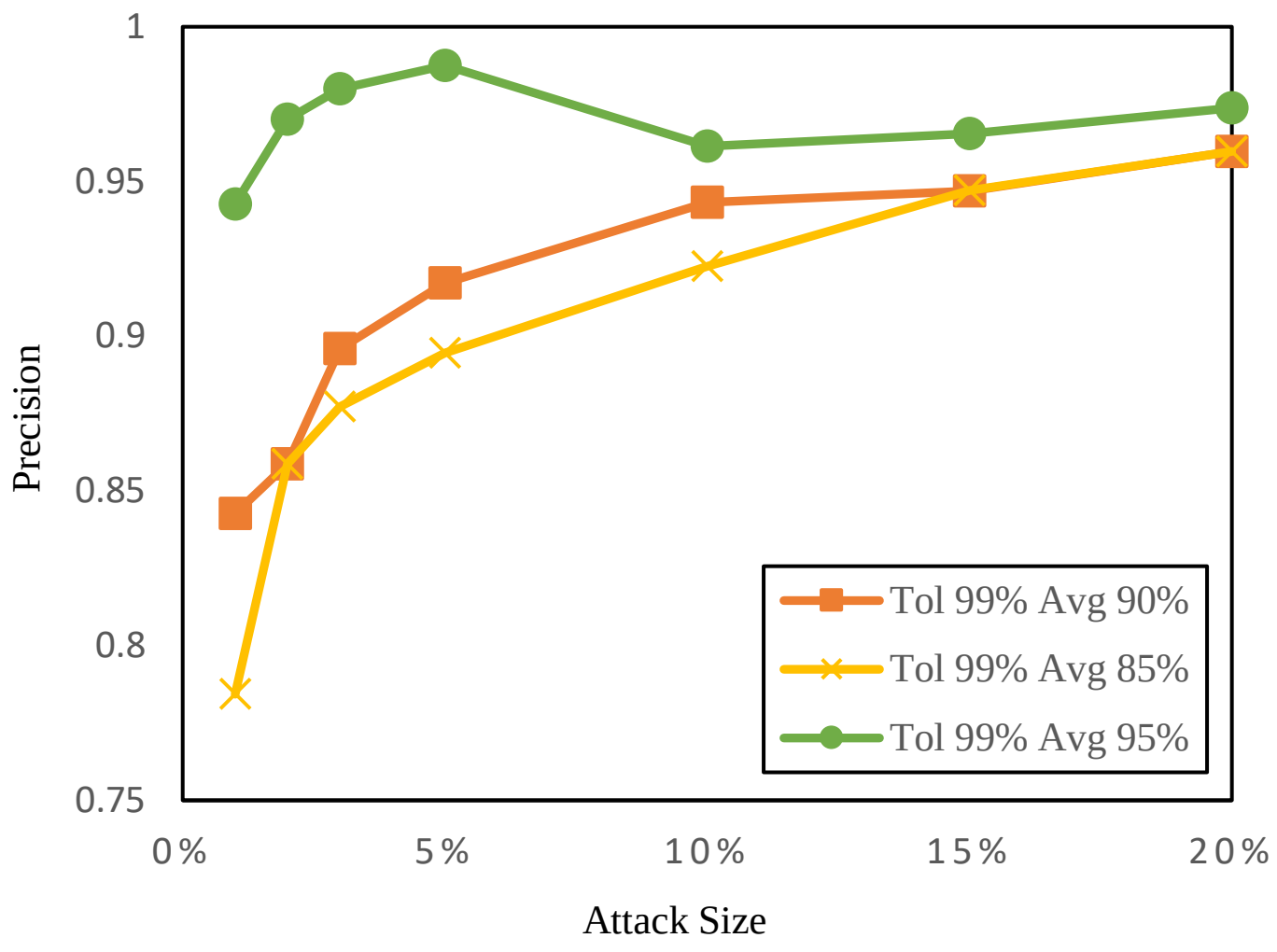

Fig. 6. The effect of different confidence levels of average deviation on detection (bandwagon attack)

Figure 7 is the recall rate of the five combinations mentioned above.

Based on the above comparison, for the confidence level of the total deviation, the higher the confidence level, the higher the precision. The recall is almost the same, so we choose a 99% confidence level.

For the confidence level of the average deviation, both the 85% and 95% confidence levels have significant disadvantages under certain circumstances. So we choose a more balanced 90% confidence level.

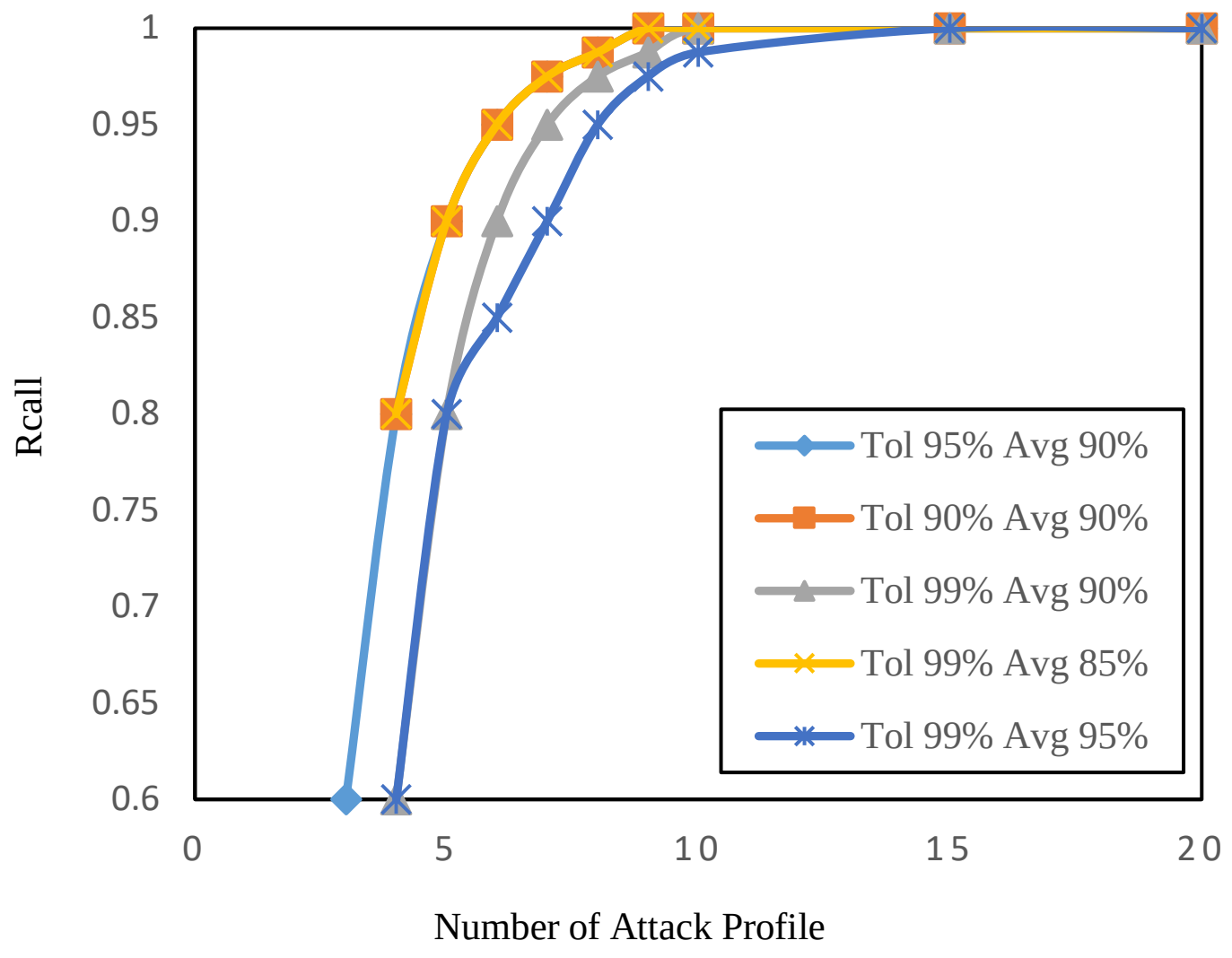

Fig. 7. The recall of different confidence levels (bandwagon attack)

*3) Comparison of detection performance*

We compared our approach with three algorithms. As shown in Figures 8, in terms of the precision of the average attack, our method is slightly better than the improved SVM algorithm (slightly less than it in the case of low attack size) and is significantly better than the other two algorithms. Bayes algorithm is not suitable for detection on low attack size.

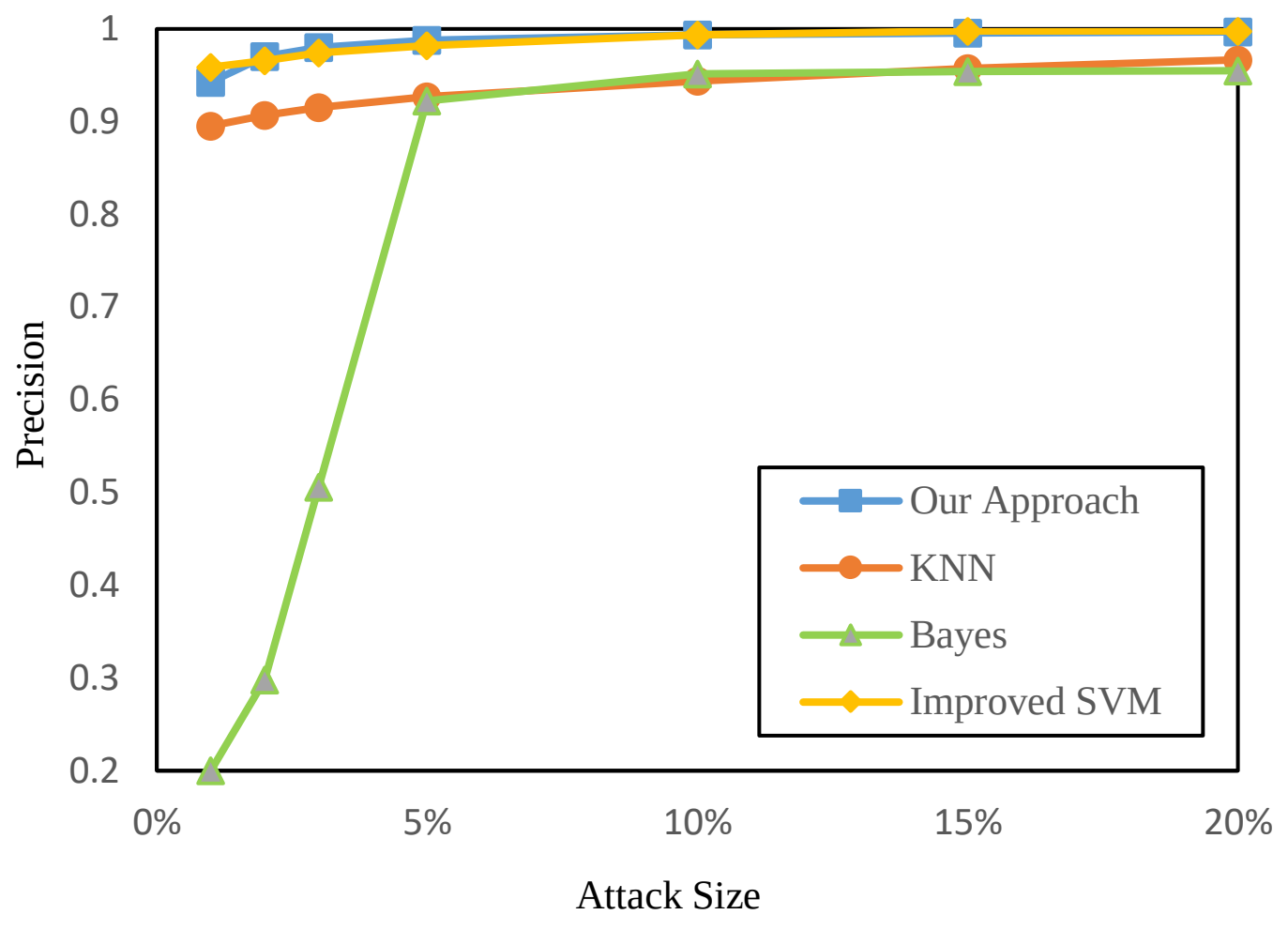

Fig. 8. Precision of different detection algorithms on average attack with a filler size of 5%

The precision contrast in the bandwagon attack is shown in the Figure 9, the precision of other algorithms does not change much compared to it in the average attack, and the precision of our algorithm has a certain decline due to the influence of the selected item.

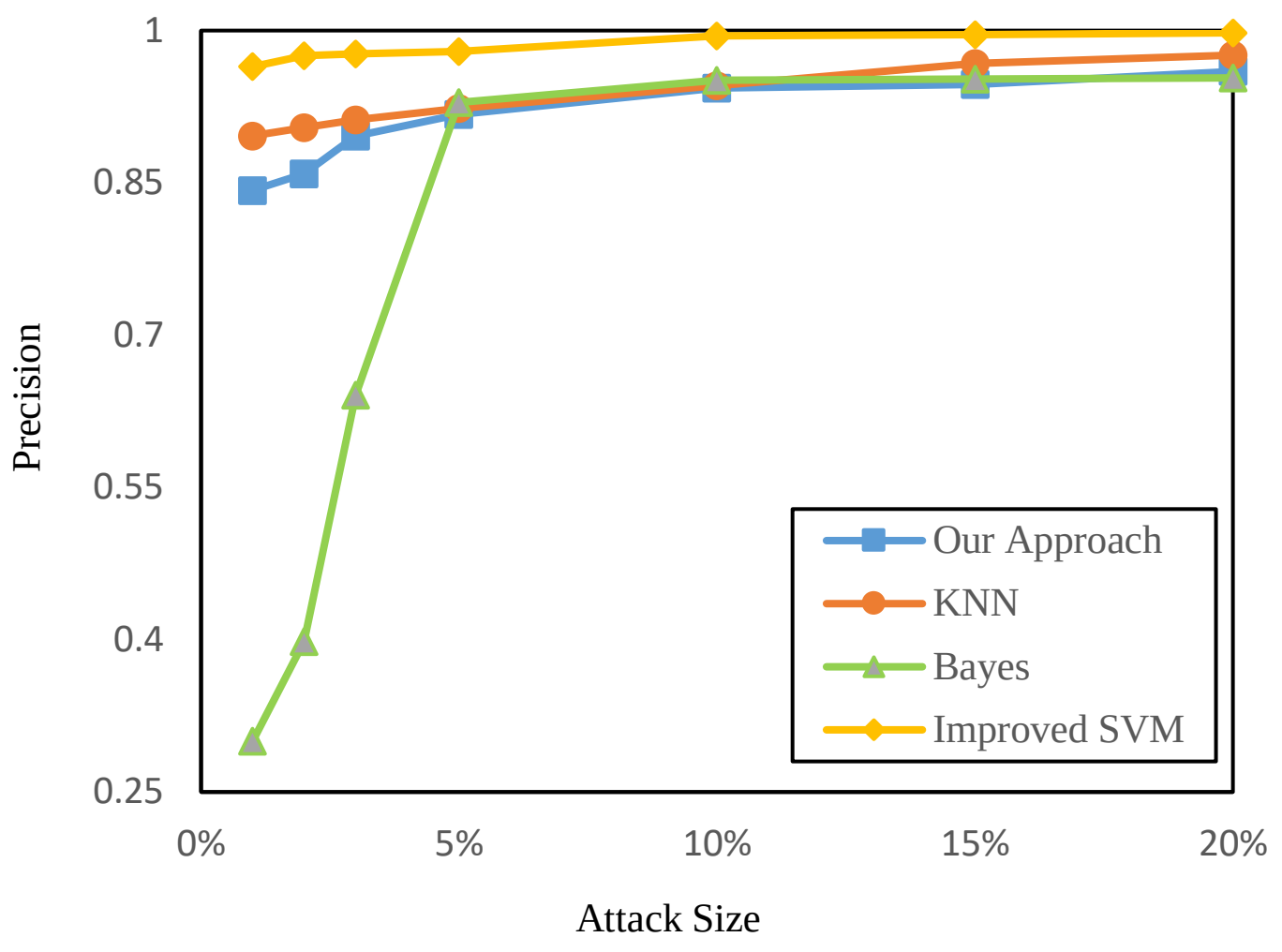

Fig. 9. Precision of different detection algorithms on bandwagon attack with a filler size of 5%

Figure 10 is the recall comparison of the three algorithms in the average attack. Our algorithm shows the best performance. The improved SVM algorithm has reached a very high precision level due to the addition of the target item analysis strategy. However, this strategy is mainly to reduce the situation that genuine users are classified as attack users. Therefore, its recall is limited to the SVM model itself does not reach a higher level. The combiner strategy used in the Bayes detection algorithm is also for misclassification. So, its recall is also based on the performance of the Bayesian model itself. The KNN algorithm responds to the assumption that the attacker has a high similarity, and can clearly distinguish between the genuine user and the attack user, but for those attack users whose attack effect is not obvious, there is no good detection effect (low recall).

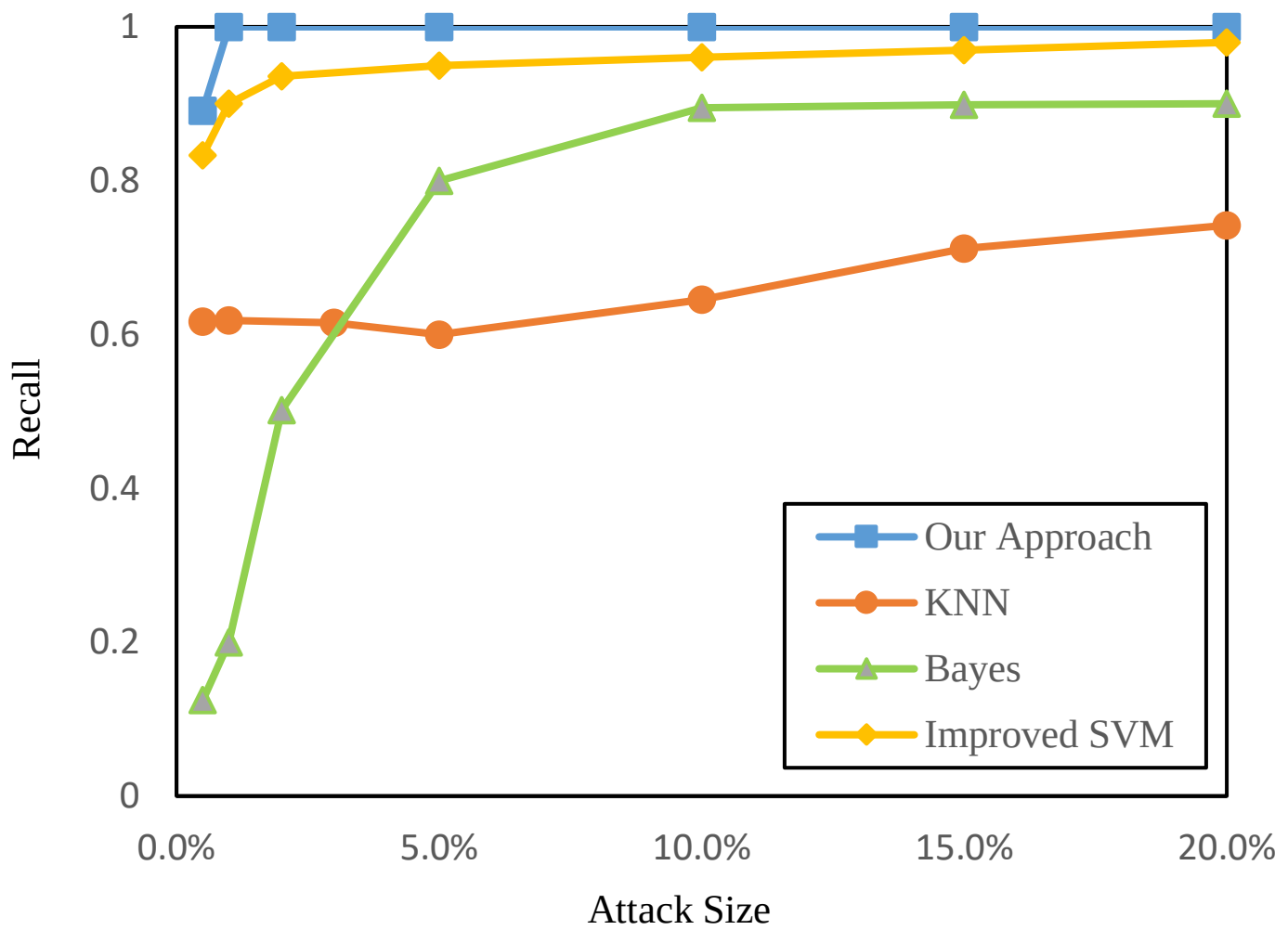

Fig. 10. Recall of different detection algorithms on average attacks with a filler size of 5%

Our method is different from other algorithms in terms of user vectors, but detects the true and false of the rating itself. It is highly sensitive to abnormal conditions, which is reflected in the high recall. Since it is targeted at rating data, the likelihood of misclassification increases when the rating is mapped to the user, that is, the precision is lower in some cases.

## VI. CONCLUSIONS AND FUTURE WORK

In this paper, we propose an improved Kalman filter model RDAKF. It calculates the deviation between the predicted value and the actual value, and measures the abnormal rating data according to the confidence interval estimation. The experiment uses the dataset of Movielens to analyze the performance of our approach under different attack models. Then, we compare our algorithm with other algorithms. The experimental results show that our algorithm has a good effect in detecting the abnormal ratings, but it is not effective for some attack models with selected items.

In the future work, we will focus on the impact of selected items and consider more types of attacks. Combine some implicit features with more datasets to come up with a better solution.

## ACKNOWLEDGMENT

This work is supported by the National Nature Science Foundation of China (61702368), Major Research Project of National Nature Science Foundation of China (91646117) and Natural Science Foundation of Tianjin (17JCYBJC15200, 18JCQNJC0070)